# New Discoveries in Galaxies across Cosmic Time through Advances in Laboratory Astrophysics

Submitted by the

American Astronomical Society Working Group on Laboratory Astrophysics
http://www.aas.org/labastro/

Nancy Brickhouse - Harvard-Smithsonian Center for Astrophysics
nbrickhouse@cfa.harvard.edu, 617-495-7438

John Cowan - University of Oklahoma
cowan@nhn.ou.edu, 405-325-3961

Paul Drake - University of Michigan
rpdrake@umich.edu, 734-763-4072

Steven Federman[*] - University of Toledo
steven.federman@utoledo.edu, 419-530-2652

Gary Ferland - University of Kentucky
gary@pa.uky.edu, 859-257-879

Adam Frank - University of Rochester
afrank@pas.rochester.edu, 585-275-1717

Eric Herbst - Ohio State University
herbst@mps.ohio-state.edu, 614-292-6951

Keith Olive - University of Minnesota
olive@physics.umn.edu, 612-624-7375

Farid Salama - NASA/Ames Research Center
Farid.Salama@nasa.gov, 650-604-3384

Daniel Wolf Savin[*] - Columbia University
savin@astro.columbia.edu, 1-212-854-4124,

Lucy Ziurys – University of Arizona
lziurys@as.arizona.edu, 520-621-6525

[*] Co-editor

# 1. Introduction

As the Galaxies across Cosmic Time (GCT) panel is fully aware, the next decade will see major advances in our understanding of these areas of research. To quote from their charge, these advances will occur in studies of "the formation, evolution, and global properties of galaxies and galaxy clusters, as well as active galactic nuclei and QSOs, mergers, star formation rate, gas accretion, and supermassive black holes."

Central to the progress in these areas are the corresponding advances in laboratory astrophysics that are required for fully realizing the GCT scientific opportunities within the decade 2010-2020. Laboratory astrophysics comprises both theoretical and experimental studies of the underlying physics that produce the observed astrophysical processes. The 5 areas of laboratory astrophysics that we have identified as relevant to the CFP panel are atomic, molecular, solid matter, plasma, nuclear, and particle physics.

In this white paper, we describe in Section 2 some of the new scientific opportunities and compelling scientific themes that will be enabled by advances in laboratory astrophysics. In Section 3, we provide the scientific context for these opportunities. Section 4 briefly discusses some of the experimental and theoretical advances in laboratory astrophysics required to realize the GCT scientific opportunities of the next decade. As requested in the Call for White Papers, Section 5 presents four central questions and one area with unusual discovery potential. Lastly, we give a short postlude in Section 6.

# 2. New scientific opportunities and compelling scientific themes

The X-ray spectra of many active galactic nuclei (AGNs) show numerous absorption features produced by ionized gas in an outflowing wind. The significance is still being debated of these winds for the AGN central engine or for the host galaxy in terms of mass, energy, and momentum. These outflows may also play a central role in cosmological feedback and in the metal enrichment of the intergalactic medium (Holczer et al. 2007).

Cooling flows in the cores of clusters of galaxies have been a subject of intense study for several decades and the details of the cooling process remain highly uncertain. *XMM-Newton* spectroscopic observations "exhibit strong emission from cool plasma at just below the ambient temperature $T_0$, down to $T_0/2$, but also exhibit a severe deficit of emission relative to the predictions of isobaric cooling-flow model at lower temperatures ($<T_0/3$). In addition, the best-resolved spectra show emission throughout the entire X-ray temperature range, but increasingly less emission at lower temperatures than the cooling-flow model would predict. These results are difficult to reconcile with simple prescriptions for distorting the emission measure distribution, e.g., by including additional heating or rapid cooling flows" (Peterson et al. 2003).

*XMM-Newton* observations of elliptical galaxies have "derived metal abundances which are higher than have usually been inferred from prior, lower spectral resolution observations [of NGC 4636] but are still incompatible with conventional chemical-enrichment models of elliptical galaxies. In addition, [these observations] are incompatible with standard cooling-flow models for this system" (Xu et al. 2003).

One of the big remaining needs for X-ray astronomy is spatially-resolved observations at high spectroscopic resolution for galaxies, clusters of galaxies, and supernova remnants (SNRs). Such observations promise to provide a wealth of new information about the properties of these objects.

Recent observations of galaxies in the distant Universe reveal the presence of interstellar material much like that seen in the Milky Way. For instance, HCN emission, a dense gas tracer, is seen in sub-millimeter galaxies and QSOs (Gao et al. 2007). Even the more widespread emission associated with PAHs is observed in distant galaxies (e.g., Rigby et al. 2008). As more detailed studies become available over the next decade, the correspondence between the emission from local gas clouds and those seen in galaxies from earlier epochs will be more clearly discerned.

## 3. Scientific context

The upcoming decade promises numerous opportunities for progress in the areas of GCT. There are a multitude of current and planned ground-based and satellite observatories including (but not limited to) the Atacama Large Millimeter/submillimeter array *(ALMA)*, *Chandra*, the Gemini telescopes, *Herschel*, the Hobby-Eberly telescope *(HET)*, *Hubble*, the Institut de RadioAstronomie Millimétrique *(IRAM)*, the James Clerk Maxwell Telescope *(JCMT)*, *JWST*, *Keck*, *Kepler*, the Large Binocular Telescope *(LBT)*, Owens Valley Radio Observatory *(OVRO)*, *Magellan*, MMT Observatory *(MMTO)*, *NEXT*, *SOFIA*, *Spitzer*, the Southern African Large Telescope *(SALT)*, the Submillimeter Array *(SMA)*, *Subaru*, the Swedish-European Space Organization Submillimeter Telescope *(SEST)*, the *Very Large Array (VLA)*, and *XMM-Newton*. Studies from the radio to X-rays contribute to our overall understanding of the physical and chemical processes important in GCT. Interpreting these data will require an accurate understanding of the underlying microphysics that produces the observed spectra.

Recent *Chandra* and *XMM-Newton* spectroscopic observations of AGNs have provided a wealth of data and discovery in these objects. From the presence of X-ray absorbing outflows to the existence of new phases of gas in these outflows as detected through absorption by Fe M-shell unresolved transition array (UTA) features, astrophysicists have deepened their understanding of these objects while simultaneously raising new questions. Answering these questions is one of the major drives behind the planned launch of the *International X-Ray Observatory (IXO)*.

X-ray spectroscopic observations for cooling-flows and metal abundances in clusters of galaxies and for elliptical galaxies are a major impetus behind future X-ray satellite observatories. Particularly important are spatially and spectrally resolved observations such as will be provided by the planned microcalorimeter on the Japanese satellite mission *NEXT*. Such observations will allow researchers to use spectroscopic diagnostics to develop a spatially resolved map for the physical properties of the cooling-flows and the metal abundances. The expected microcalorimeter observations of galaxies and SNRs also are expected to revolutionize our understanding of these sources.

In our Galaxy, dense molecular clouds are the sites of star formation. They appear to be involved with star formation in the Galactic Center, even in the presence of phenomena associated with a supermassive black hole. Moreover, PAH emission is ubiquitous in clouds within the disk of our Galaxy. Now that molecular emission from dense clouds and PAH emission are seen is distant galaxies, it is important to examine whether the inferred properties derived locally apply to galaxies at a much earlier cosmic time. The new facilities nearing completion will allow more detailed comparisons than ever before.

## 4. Required advances in Laboratory Astrophysics

Advances particularly in the areas of atomic, molecular, solid matter, plasma, nuclear, and particle physics will be required for the scientific opportunities described above. Here we briefly discuss some of the relevant research in each of these 6 areas of laboratory astrophysics. Experimental and theoretical advances are required in all these areas for fully realizing the GCT scientific opportunities of the next decade.

### 4.1. Atomic Physics

Reliably characterizing the speed of the AGN outflows requires accurate experimental and theoretical wavelengths for the X-ray absorption lines used to measure the outflow velocities.

The shapes, central wavelengths, and equivalent widths of the AGN UTAs can be used to diagnose the properties of the AGN warm absorbers (Behar et al. 2001). However models that match K-shell absorption features from 2nd and 3rd-row elements cannot reproduce correctly the observed UTAs from the fourth-row element iron (Netzer et al. 2003). The problem is due in part to the lack of reliable low temperature dielectronic recombination data for Fe M-shell ions (Netzer 2004, Kraemer et al. 2004).

Spectroscopic diagnostics using lines from He-like ions and line ratios from other isoelectronic sequences can give temperature measurements independent of ionization state. Because these results do not require a model for the charge state distribution of the gas, they are robust for both ionization equilibrium and more importantly for cases of non-equilibrium ionization. For clusters of galaxies, it is known that resonance scattering of hydrogenic-like Fe Lyman $\alpha$ and other resonance series lines is important. Hence, given the inferred temperature, the Lyman $\alpha$ and other resonance series line ratios can be used to derive an independent measure of the column density for clusters. Accurate theoretical line ratios are crucial for these temperature measurements. Uncertainties on the order of 20% can result in a factor of 2 error in the inferred temperature (Chen et al. 2006).

Reliably determining the electron and ion temperatures, densities, emission measure distributions, and ion and elemental abundances of cosmic atomic plasmas requires accurate fractional abundance calculations for the different ionization stages of the various elements in the plasma (i.e., the ionization balance of the gas). Since many of the observed sources are not in local thermodynamic equilibrium, in order to determine the ionization balance of the plasma one needs to know the rate coefficients for all of the relevant recombination and ionization processes. Radiative recombination, dielectronic recombination, and charge transfer are the primary recombination processes to consider. Ionization includes that due to photons, electrons, and from charge transfer. Considering all the ions and levels that need to be taken into account, it is clear that vast quantities of data are needed. Generating them to the accuracy required for astrophysics pushes theoretical and experimental methods to the edge of what is currently achievable and often beyond. For this reason progress is slow and every 5-10 years a group of researchers takes it upon themselves to survey the field, evaluate and compile the relevant rate coefficients, and provide the most reliable ionization balance calculations possible at that time (Savin 2005, Bryans et al. 2006). Such papers are among the most highly cited laboratory astrophysics publications (e.g., Mazzotta et al. 1998 with over 500 citations).

Analyzing and modeling spectra begin with identifying the observed lines that may be seen in emission or absorption. This requires accurate and complete wavelengths across the electromagnetic spectrum. The next step toward understanding the properties of an observed cosmic source depends on accurate knowledge of the underlying atomic processes producing the observed lines. Oscillator strengths and transition probabilities are critical to a wide variety of temperature and abundance studies from infrared to X-ray wavelengths. Many existing data for the heavier elements are still notoriously unreliable.

Line ratios, a key diagnostic, involve knowledge of collision strengths and related phenomena. Rate coefficients for electron impact excitation approaching 10% accuracy are necessary for the most important line ratio diagnostics yielding temperature, optical depth, density, and abundance. Proton impact excitation is important because ions in hot post-shock material decouple from radiatively cooling electrons and may remain hot enough to produce line emission through collisional impact. Furthermore, electron impact ionization (EII) data are highly suspect. Recommended data derived from the same scant set of measurements and calculations can differ by factors of 2 to 3. Much of the published experimental data include contributions from an unknown metastable fraction in the ion beams used. Little data also exist for three-body recombination, the time reverse of EII, which is important in high density plasmas.

### 4.2. Molecular Physics

High-resolution laboratory spectroscopy is absolutely essential in establishing the identity and abundances of molecules observed in astronomical data. Given the advancements in detector technologies, laboratory measurements need to have a resolving power higher than the astronomical instruments. This is essential to interpret ubiquitous interstellar spectral features such as the IR emission bands ("PAH bands") which hold the key to our understanding of the molecular universe (Salama 1999). It is also crucial in making the link between interstellar molecules in the gas phase and dust grains in the solid phase, a key phase in our understanding of the evolution of interstellar clouds. For molecular data obtained from astronomical observations to be of practical use, accurate assignments of observed spectral features are essential. The problem here is two-fold. First, the transitions of known molecules need to be assigned in these spectra, including higher energy levels and new isotopic species. Second, the spectra of undiscovered species that promise to serve as important new probes of astronomical sources need to be identified.

The spectroscopic study of interstellar molecules, many of which are complex structures that cannot be produced in large abundance in the laboratory, requires the development and application of state-of-the-art ultra-sensitive spectroscopic instruments. Detecting the possible presence of a species, however, is not sufficient since it must be reconciled with other physical properties of the medium. To understand the chemical composition of these environments and to direct future molecular searches in the framework of future astronomical observatories, it is important to untangle the detailed chemical reactions and processes leading to the formation of molecules in extraterrestrial environments. These data, together with quantum chemical calculations, will establish credible chemical models.

### 4.3. Solid Material

In order to properly decipher the mechanisms that occur in interstellar cloud environments, laboratory studies of dust and ices are required (Tielens 2005). Studies of silicate and carbonaceous dust precursor molecules and grains are needed as are studies of the dynamical interaction between dust and its environment (including radiation and gas). Astronomical observations and supporting laboratory experiments from the X-ray domain to the sub-millimeter regions are of paramount importance for studies of the molecular and dusty universe. Observations at infrared and sub-millimeter wavelengths penetrate the dusty regions and probe the processes occurring deep within them. These wavelengths provide detailed profiles of molecular transitions associated with dust.

Mid-IR spectra of individual objects such as H II regions, reflection nebulae, and planetary nebulae as well as the general interstellar medium are dominated by a set of emission features due to large aromatic molecules (a.k.a "PAH bands"). Studies of the spectral characteristics of such molecules and their dependence on molecular structure and charge state are of key importance for our understanding of this ubiquitous molecular component of the ISM. At long wavelengths, the continuum dust opacity is uncertain by an order of magnitude. IR spectral features of interstellar dust grains are used to determine their specific mineral composition, hence their opacities, which determine inferred grain temperatures and the masses of dusty objects, including the interstellar medium of entire galaxies. Emission bands from warm astronomical environments such as star-forming clouds lead to the determination of the composition and physical conditions in these regions. The laboratory data essential for investigations of dust include measurements of the optical properties of candidate grain materials (including carbonaceous and silicate materials, as well as metallic carbides, sulfides, and oxides) as a function of temperature. For abundant materials (e.g., forms of carbon such as PAHs), the measurements should range from gas-phase molecules to nanoparticles to bulk materials. The IR spectral region is critical for the identification of grain composition, but results are also required for shorter wavelengths (i.e., UV), which heat the grains. The lack of experimental data in this spectral region has hampered progress in theoretical studies as well as the interpretation of astronomical data.

### 4.4. Plasma Physics

Plasma physics at high energy density can produce photoionized plasmas that can benchmark models of AGN and QSO absorbers. This is accomplished by producing an intense x-ray source that can irradiate a plasma volume containing relevant species, and measuring the ionization balance and other properties that result. Work in this direction has begun (Foord et al. 2006; Wang et al. 2008). Much more will be possible in the coming decade as higher-energy facilities come online. Another contribution is to produce environments with intense magnetic field such as exist in the vicinity of AGNs and QSOs. Understanding the behavior of charged particles in intense magnetic fields is important for star formation, gas accretion, feedback, and supermassive black holes. Such fields can also alter transition energies affecting the analysis of spectra from the vicinity of highly magnetized plasma. Gigagauss fields, large enough to have significant effects, are produced by current-generation intense lasers (Wagner et al. 2004); the next generation will reach 10 Ggauss.

### 4.5 Nuclear Physics

Experimental nuclear lab cross sections are helping us to determine heavy element abundances, particularly those from the r-process. Those abundance patterns in Galactic stars are "solar like", indicating surprisingly that the same types of nuclear processes have been occurring over Gyr in our Galaxy. Even more surprising observational studies of stars in distant galaxies, in for example Damped Lyman Alpha Systems, and in nearby galaxies also show the same solar system pattern for heavy elements. This points to a commonality in nucleosynthesis in the earliest stars in other galaxies (Cowan 2007).

A recent, much improved measurement at the underground LUNA facility, Gran Sasso, Italy, of the cross section for $^{14}N(p,\gamma)$ has altered the age estimates of globular clusters, increasing them by about 0.9 Gyr. Further improvements are anticipated. This reaction sets the clock for the CN cycle.

### 4.6 Particle physics

Galaxy formation rests on an initial set of seeds which represent a departure from homogeneity in the Universe. The amplitude of these perturbations is constrained by the observations of power spectrum in the cosmic microwave background (CMB) radiation. It is well known that these perturbations (thought to have been produced in the very early Universe during inflation) would not have sufficient time to grow in a baryon dominated universe. The accepted solution to this problem is dark matter. If dark matter is not coupled to radiation, then the perturbations in the dark matter can begin to grow under the influence of gravity before the epoch of the last scattering of the CMB. As a consequence, when atoms finally become neutral and are free to collapse into large scale structures, they already see well formed potentials in dark matter. In this way, galaxies and clusters of galaxies are seeded by existing structure in the dark matter. The study of dark matter in a multitude of laboratory experiments will help shed light on the important process of galaxy formation.

## 5. Four central questions and one area with unusual discovery potential

### 5.1 Four central questions

- What is the nature of the AGN warm absorber outflows?
- What is the correct physical description for cooling flows in galaxy clusters?

- What are the metal abundances and cooling-flow properties for elliptical galaxies?
- What is the correspondence between star formation and activity in our Galaxy and what is seen in the distant Universe?

### 5.2 One area with unusual discovery potential

- New facilities will provide the data needed to translate knowledge gained on our Galaxy and its neighbors with an understanding of the distant Universe.

### 6. Postlude

Laboratory astrophysics and complementary theoretical calculations are part of the foundation for our understanding of GCT and will remain so for many generations to come. From the level of scientific conception to that of the scientific return, it is our understanding of the underlying processes that allows us to address fundamental questions in these areas. In this regard, laboratory astrophysics is much like detector and instrument development; these efforts are necessary to maximize the scientific return from astronomical observations.

### References

Behar, E., et al. 2001, Astrophys. J., 563, 504.
Bryans, P., et al. 2006, Astrophys. J. Suppl. Ser., 167, 343.
Chen, G. X., et al. 2006, Phys. Rev. A, 74, 042709.
Cowan, J. 2007, Nature, 448, 29.
Foord, M. E., et al. 2006, J. Quant. Spectrosc. Radiat. Transfer, 99, 712.
Gao, Y., et al. 2007, Astrophys. J., 660, L93.
Holczer, T., et al. 2007, Astrophys. J., 663, 799.
Kraemer, S., et al., 2004, Astrophys. J., 604, 556.
Mazzotta, P., et al. 1998, Astron. Astrophys. 133, 403.
Netzer, H., et al., 2003, Astrophys. J., 599, 933.
Netzer, H. 2004, Astrophys. J., 604, 551.
Peterson, J. R., et al. 2003, Astrophys. J., 590, 224.
Rigby, J. R. et al. 2008, Astrophys. J., 675, 262.
Salama, F. 1999 in Solid Interstellar Matter: The ISO Revolution, d'Hendecourt et al. eds., Savin, D. W. 2005, AIP Conf. Proc., 774, 297.
Tielens, A. G. G. M. 2005, The Physics and Chemistry of the Interstellar Medium, Cambridge University Press.
Wagner, U., et al. 2004, Phys. Rev. E, 70, 026401.
Wang, F. L., et al. 2008, Phys. Plasmas, 15, 073108.
Xu, H., et al. 2003, Astrophys. J., 579, 600.